\documentstyle[sprocl]{article}
\input {epsf}
\def\plotone#1{\centering \leavevmode
\epsfxsize= 0.4\columnwidth \epsfbox{#1}}
\def\plottwo#1#2{\centering \leavevmode
\epsfxsize=.43\columnwidth \epsfbox{#1} \hfil
\epsfxsize=.43\columnwidth \epsfbox{#2}}

\newdimen\hhsize\hhsize=.5\hsize

 \def\PRL{{\em Phys. Rev. Lett.}}
\def\PRD{{\em Phys. Rev.} D}  

\def\be{\begin{equation}} \def\ee{\end{equation}}
\def\bea{\begin{eqnarray}} \def\eea{\end{eqnarray}}
 \def\be{\begin{equation}}
\def\ee{\end{equation}} \def\bea{\begin{eqnarray}}
\def\eea{\end{eqnarray}}

\def\simgreat{\mathrel{\mathpalette\fun >}} 

\def\GeV{\,{\rm GeV}}

\def\cmm2{{\,\rm cm^{-2}}}
\def\cm2{{\,{\rm cm}^2}}
\def\cmm3{{\,{\rm cm}^{-3}}}
\def\gcmm3{{\,{\rm g\,cm^{-3}}}}

\def\R{{\cal R}}

\def\fun#1#2{\lower3.6pt\vbox{\baselineskip0pt\lineskip.9pt
  \ialign{$\mathsurround=0pt#1\hfil##\hfil$\crcr#2\crcr\sim\crcr}}}

\def\tlh{\partial\ln H/\partial\phi}
\def\tlhh{\partial^2\ln H/\partial\phi^2}

\def\tmpls{ (m^2_{\cal P}/4\pi)}
\def\today{\ifcase\month\or
 January\or February\or March\or April\or May\or June\or
 July\or August\or September\or October\or November\or December\fi
 \space\number\day, \number\year}

\begin{document}
\title{
Prospects for measuring  Inflation parameters with the CMB
}
\author{{\bf Tarun Souradeep}$^{1}$, J. Richard Bond$^{1}$, Lloyd Knox$^{1}$, 
\\
George Efstathiou$^{2}$ and  Michael S. Turner$^{3}$ }
\address{1. CITA, University of Toronto, Toronto, ON M5S 3H8, CANADA \\
2.    IOA, University of Cambridge, Cambridge, CB3 0HA, UK \\
3. Department of Physics \& Astrophysics, Enrico Fermi Institute,\\
University of Chicago, Chicago, IL 60637-1433}

\maketitle\abstracts{ We present the status of our study into the
feasibility of estimating the parameters of inflation from planned
satellite observations of the anisotropy of the cosmic microwave
background (CMB).We describe a perturbative procedure for studying
inflation models with soft departures from scale free spectra and find
that in this case the parameters of inflation are likely to be well
determined. We also demonstrate that radical departures from scale
free initial perturbations with broad features in $k$-space can be be
measured by Planck. However, allowing for very fine features in
$k$-space can very strongly compromise the ability to measure
cosmological parameters.}

\vskip -0.15in
\noindent{\bf 1. Introduction}

\noindent Planned satellite observations of the anisotropy of the
cosmic microwave background (CMB) are expected to determine the
parameters associated with cosmological structure formation to
unprecedented accuracy ~\cite{paramest,bet}.  Inflation remains the
best motivated and completely predictive early universe scenario that
is invoked to specify the spectrum of initial perturbations for
structure formation.  In this article, we report on our investigation
into the prospects of measuring the parameters of inflationary models
using the high resolution, full sky CMB anisotropy data expected in
the near future from the MAP and Planck Surveyor satellite
missions. Based on the `Hamilton-Jacobi formulation', we parameterize
the inflationary phase by the `action', $H(\phi)$ -- a function of the
inflaton field $\phi$. We then seek to determine how well one can
constrain $H(\phi)$ using MAP and Planck data. For a single inflaton
field, the determination of $H(\phi)$ can be directly translated to an
estimation of the inflaton potential,$V(\phi) = (3 m^2_{\cal P}/8\pi)
H^2(\phi) [1 - (m^2_{\cal P}/12\pi)(\tlh)^2 ] $, using the `reduced
Hamilton-Jacobi equation'~\cite{bon_yuk}. The inflaton potential
reconstructed from the CMB anisotropy data will herald a new probe of
physics at ultra-high energy scales ($\simgreat 10^{14}
\GeV$)~\cite{l2kcba97}.

We assume that, given the MAP or Planck data, a set of model
parameters $\bf{y}$ will be varied to find the $\bf{y_*}$ that
minimizes log of the likelihood, $\ln{\cal L}$. The $\bf{y_*}$ is then
the target model around which we estimate the errors on the parameter.
The error matrix at $\bf{y_*}$ is given by $(C^P)^{-1}_{ij}=
-\partial^2\langle\ln{\cal L}\rangle/ \partial y_i \partial y_j$,
where $C^P$ is the parameter covariance matrix. We include a set of
parameters specifying the initial perturbations from inflation, in
addition to the set of cosmological parameters.( For the choice of
cosmological parameters, specifications for the MAP and Planck
missions used and the basic methodology, see reference~\cite{bet}.)
Parameter estimation in two disparate classes of inflation models is
discussed in the next two sections.

\smallskip
\noindent{\bf 2.  Soft deviations from scale free spectra}

\noindent The simplest consideration of inflation corresponds to the
uniform acceleration (constant $\tlh$)
approximation~\cite{bon_yuk}. The initial scalar and tensor
perturbation spectra are scale free (power law) and are parameterized
by two numbers~: the scalar perturbation amplitude and a common
spectral index, $\nu = -2 \tmpls(\tlh)^2/[1-\tmpls(\tlh)^2]$.

\begin{figure}[bthp]
\plottwo{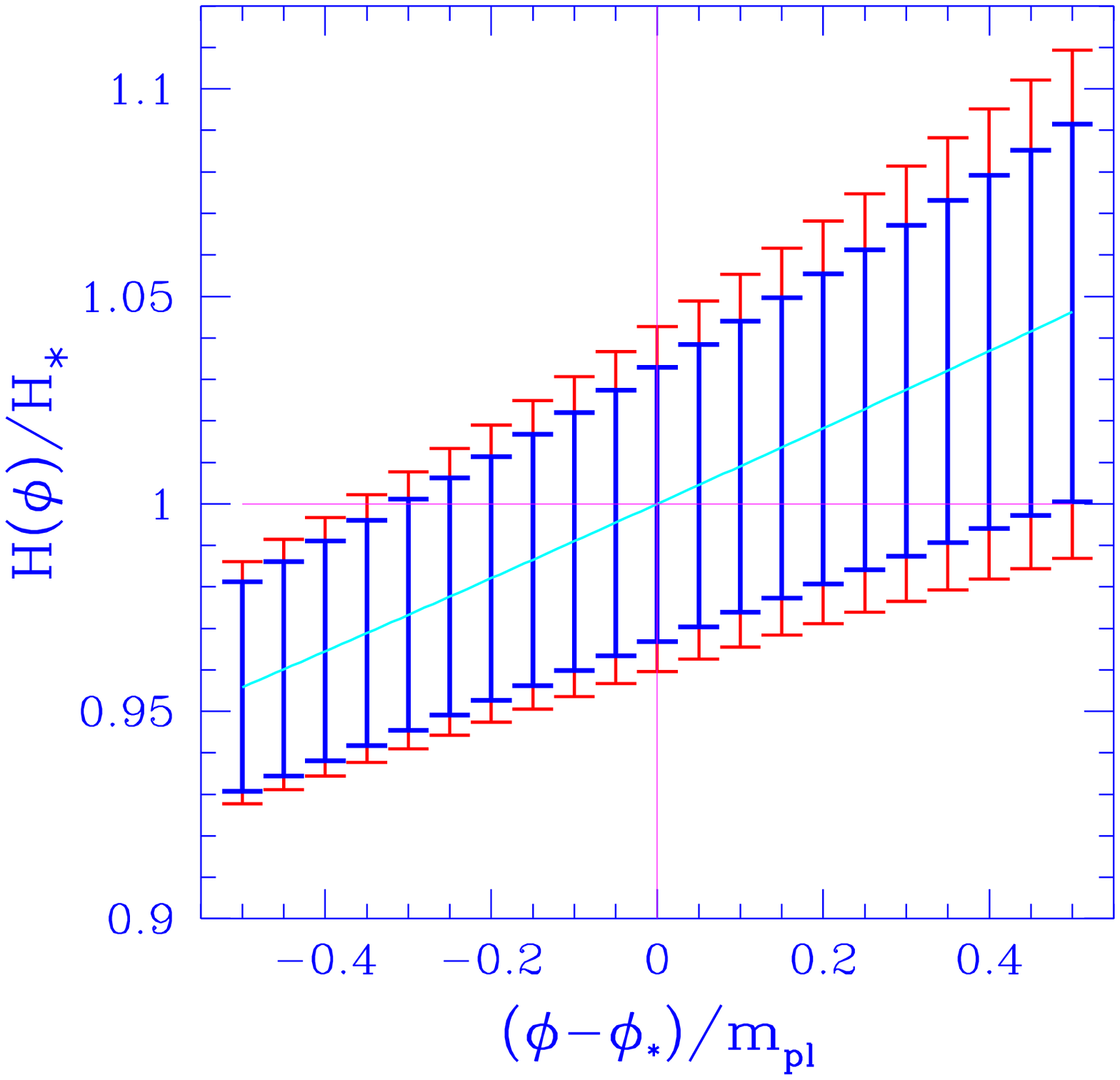}
{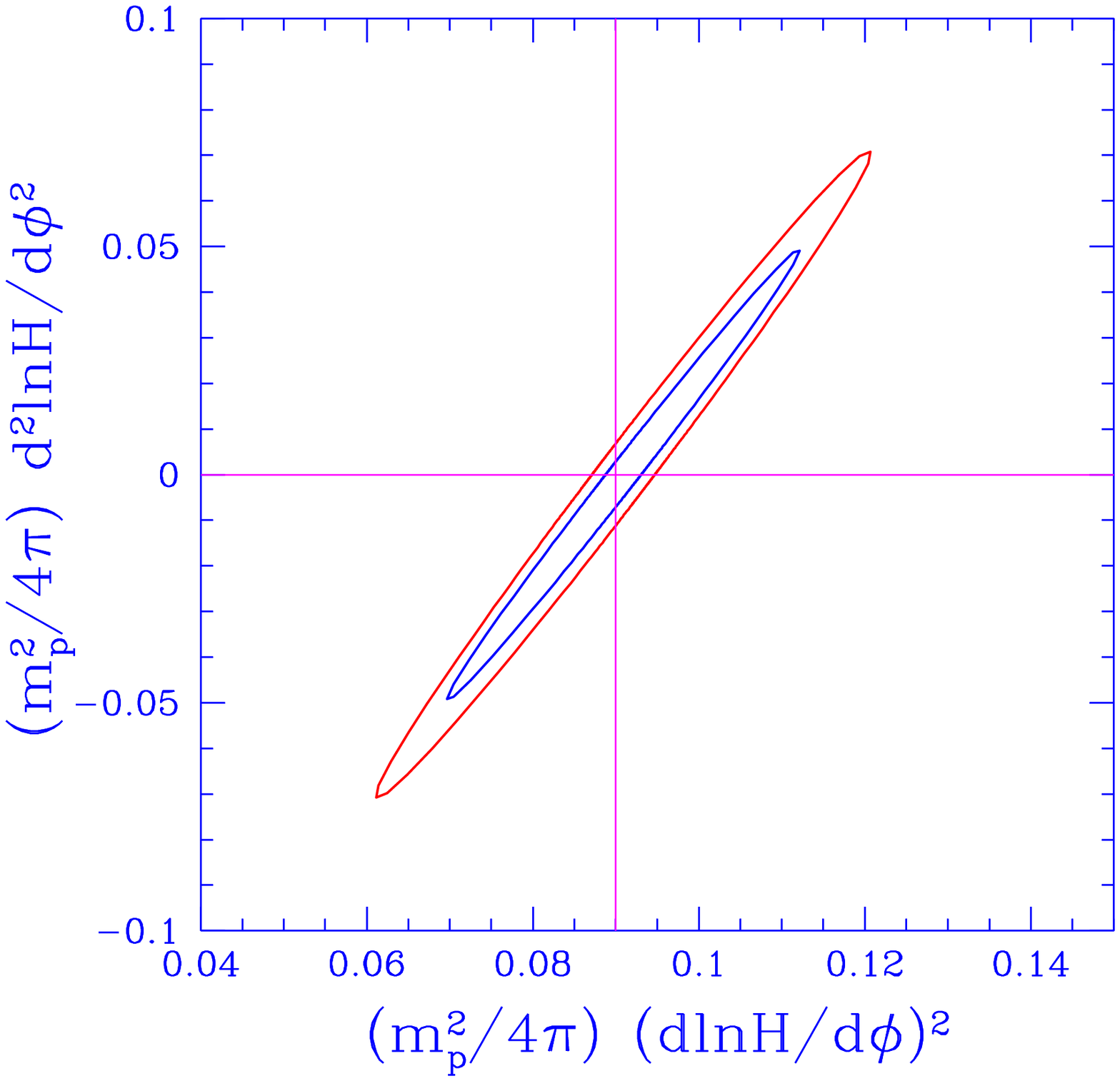}
\vspace{-0.18in}
\caption[] {$H(\phi)$ from the satellite experiments will be
determined within the $1$-$\sigma$ band shown on the {\bf left panel}.
Though Planck is a little better than MAP, the disappointing aspect is
that the amplitude is not well determined leading to a large
uncertainity in $H(\phi)$ .  As shown in the panel on the {\bf right},
the shape of $H(\phi)$ specified in terms of its derivatives is much
better determined. The {\bf right panel} shows the expected
$1$-$\sigma$ error ellipses in the $(d\ln H/d\phi)^2$--$d^2\ln
H/d\phi^2$ plane. In the above figures, larger uncertainity case
corresponds to the MAP satellite experiment.  The target model is
tilted ( $\nu_s=n_s-1= -0.2$) sCDM.}
\label{fig_one}
\end{figure}

The CMB anisotropy data from Planck and MAP would be able to measure
quite accurately deviations from power law spectra. We study models
with small deviations from uniform acceleration in terms of a small
$\tmpls\tlhh$ perturbation~\cite{us_inprep}. The predicted scalar and
tensor spectra are perturbed from the scale free form, but, at the
leading order, maintain a constant difference $\nu_{st}=
2\tmpls\tlhh/[1-\tmpls(\tlh)^2]$ between their spectral indices. It is
possible to solve for $H(\phi)$ in an exact form and consequently its
$\phi$-derivatives in terms of $\nu$ and
$\nu_{st}$. Figure~\ref{fig_one} shows the expected errors on the
inflationary parameters for a tilted sCDM target model. More details
of our results will be presented in a forthcoming
publication~\cite{us_inprep}.

\smallskip
\noindent{\bf 3. Radical departures from scale free spectra}

\noindent Even single scalar field models of inflation can produce a
wide variety of initial power spectra with radical departure from a
scale free form. In this case, connection of the estimated initial
perturbation spectra to $H(\phi)$ can be made only in a limited
sense. Here we investigate the impact on cosmological parameter
determination of radical departures from scale-free initial
conditions.

\begin{figure}[bthp]
\plotone{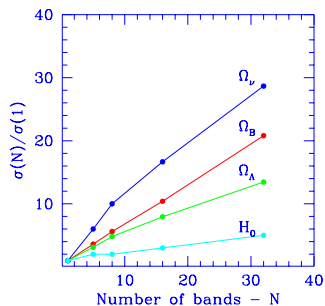}
\vspace{-0.18in}
\caption[] {The figure shows the relative increase in the expected
$1-\sigma$ error bars on selected cosmological parameters as one
increases the number of $k$-space bands.}

\label{cosm_band_var}
\end{figure}

We parameterize the initial power spectrum by specifying the amplitude
in $N$ bands in $k$-space. We enforce smoothness by using smooth shape
functions peaked at each $k$-space band. As one increases the freedom
in the power spectrum by increasing $N$, the shape functions can more
readily mimic the effect of various cosmological parameters, leading
to a gradual build up of degeneracy amongst the shape parameters and
cosmological parameters. This is reflected in increasing uncertainity
for the cosmological parameters shown in Fig.~\ref{cosm_band_var}.
This gives an idea of the freedom one can allow in the initial
perturbation spectra at the expense of increasing inability to pin
down the cosmological parameters.

\end{document}